# On Nash Equilibrium and Evolutionarily Stable States That Are Not Characterised by the Folk Theorem

Jiawei Li[1]*, Graham Kendall[1,2]

**1** ASAP Research Group, School of Computer Science, University of Nottingham, Jubilee Campus, Wollaton Road, Nottingham, NG8 1BB, United Kingdom, **2** School of Computer Science, University of Nottingham Malaysia Campus, Jalan Broga, 43500 Semenyih, Selangor Darul Ehsan, Malaysia

* jiawei.li@nottingham.ac.uk

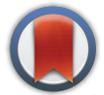

**Data Availability Statement:** All relevant data are within the paper.

**Funding:** This work was funded by Engineering and Physical Sciences Research Council UK EP/ H000968/1 (http://gow.epsrc.ac.uk/NGBOViewGrant. aspx?GrantRef=EP/H000968/1). The funders had no role in study design, data collection and analysis, decision to publish, or preparation of the manuscript.

**Competing Interests:** The authors have declared that no competing interests exist.

## Abstract

In evolutionary game theory, evolutionarily stable states are characterised by the folk theorem because exact solutions to the replicator equation are difficult to obtain. It is generally assumed that the folk theorem, which is the fundamental theory for non-cooperative games, defines all Nash equilibria in infinitely repeated games. Here, we prove that Nash equilibria that are not characterised by the folk theorem do exist. By adopting specific reactive strategies, a group of players can be better off by coordinating their actions in repeated games. We call it a type-$k$ equilibrium when a group of $k$ players coordinate their actions and they have no incentive to deviate from their strategies simultaneously. The existence and stability of the type-$k$ equilibrium in general games is discussed. This study shows that the sets of Nash equilibria and evolutionarily stable states have greater cardinality than classic game theory has predicted in many repeated games.

## Introduction

A population is considered to be in an evolutionarily stable state if its genetic composition is restored by selection after a disturbance [1]. In evolutionary game theory, an evolutionarily stable state has a close relationship with the concept of Nash equilibrium (NE) and the folk theorem for infinitely repeated games [2, 3]. The folk theorem, which is the fundamental theory of non-cooperative repeated games, states that any feasible payoff profile that strictly dominates the minimax profile is a Nash equilibrium profile in an infinitely repeated game [4, 5]. An evolutionary stable state must be a refinement of NE in the corresponding evolutionary game.

The folk theorem has been intensively studied for decades. Different variants of it have been developed to take into consideration the factors such as indefinite iteration, incomplete information and discount rate [6–15]. It is generally assumed that the folk theorem characterises all NE in an infinitely repeated game. However, there does exist some NE that are neglected by classical game theory, as we show in this paper. Let's first see a new variant of the prisoner's dilemma (PD).

We extend PD to a three-player zero-sum game by adding an extra player, the police, whose payoff is equivalent to the negative sum of the payoffs of two prisoners. Suppose, for simplicity,

If player **Z** chooses *L*:

PLAYER **Y**

|  |  | *D* | *C* |
|---|---|---|---|
| PLAYER **X** | *D* | -3, -3, 6 | 1, -5, 4 |
|  | *C* | -5, 1, 4 | 0, 0, 0 |

If player **Z** chooses *R*:

PLAYER **Y**

|  |  | *D* | *C* |
|---|---|---|---|
| PLAYER **X** | *D* | -2, -2, 4 | 2, -4, 2 |
|  | *C* | -4, 2, 2 | 1, 1, -2 |

**Fig 1. Payoff matrix of a three-player zero-sum game shows the payoffs of three players, X, Y and Z. X and Y choose between two options, *D* and *C*, whilst Z chooses between *L* and *R*.**

that the police choose between two options, *L* and *R*, which lead to two PDs between X and Y with different payoff values as shown in Fig 1.

The dominant strategy for three players are *D*, *D*, and *L* respectively and the corresponding payoffs are (-3, -3, 6), which is the unique NE of the stage game.

In an infinitely repeated version of this game, X and Y can be better off by choosing (*C*, *C*) whatever Z chooses. Three players choosing (*C*, *C*, *L*) in every round should be a NE since mutual cooperation is a NE in each PD according to the folk theorem. Note that the minimax payoffs for three players are -3, -3 and 0 respectively. Z receives the minimax payoff in this equilibrium, which means that the payoff profile does not strictly dominate the minimax payoff profile. Thus, this equilibrium is not characterised by the Folk theorem.

The strategies for players in a repeated game include not only simply aggregations of pure or mixed strategies in a sequence of stage games, but also reactive strategies that one player chooses their action in response to some other players' previous actions. From Tit for tat, Grim trigger, Pavlov and Group strategies [16] to the newly appeared zero-determinant strategies [17–19], a number of reactive strategies have been developed and investigated in evolutionary game theory. Reactive strategies are the reason why a payoff profile that is not NE in the stage game can be NE in an infinitely repeated game.

Coordination among a group of players can be formed and maintained when specific reactive strategies are adopted by those players, which leads to equilibrium that does not exist in one-shot games.

## Methods and Results

### Reactive Strategies

Consider a repeated *n*-player game $G = \{I, S, U\}^T$ where $I = \{1, \ldots, n\}$ is the player set and $S = \{S_1, \ldots, S_n\}$ and $U = \{U_1, \ldots, U_n\}$ are the strategy set and the payoff set respectively. The iteration of game is counted by *t*, starting from $t = 0$. Each player has a pure action space $A_i^t$ in the $t^{th}$ stage game. Let $\hbar_i^{\ t} = (a_i^0, a_i^1, \ldots, a_i^{t-1})$ be the sequence of actions chosen by player $i \in I$ within $t-1$ periods, and $\hbar_{-i}^t = (h_1^t, \ldots, h_{i-1}^t, h_{i+1}^t, \ldots, h_n^t)$ the past choices made by all players other than *i*. For simplicity of expression, the payoff of a player in a repeated game is computed by $u_i = \frac{1}{T+1} \sum_{t=0}^{T} u_i^t$, which denotes the average payoff over a period of *T*+1.

A player's strategy is reactive if it is a function of other players' past actions [18]. Player *i's* strategy, $s_i$, is a reactive strategy when there is

$$s_i^t = \begin{cases} s_i^0 & t = 0 \\ f(\hbar_{-i}^{\ t}) & t \geq 1 \end{cases} \qquad (1)$$

The strategy in the first stage game, $s_i^0$, is either a pure strategy or a mixed strategy. Obviously, reactive strategies do not exist in one-shot games since there always are $\hbar_i^t = \hbar_{-i}^t = \phi$ for any $i$.

Reactive strategies provide a way of coordination among a group of players in repeated games. In a repeated game with multiple Nash equilibria, for example, convergence to a Nash equilibrium can be guaranteed only if the players adopt specific reactive strategies.

There are two pure-strategy NEs, $(L, R)$ and $(R, L)$, in the coordination game as shown in Fig 2. Two players do not have any a priori knowledge about which NE strategy profile to choose unless they can communicate with each other before the game. The coordination between X and Y can be achieved with probability $\rho \rightarrow 1$ in an infinite repeated game if two players adopt the below strategies:

$$s_{Row}^t = \begin{cases} L & \text{if both players chose } (L, R) \text{ at } t - 1 \\ R & \text{if both players chose } (R, L) \text{ at } t - 1 \\ rand\{L, R\} & \text{Otherwise} \end{cases} \quad (2)$$

$$s_{Col}^t = \begin{cases} R & \text{if both players chose } (L, R) \text{ at } t - 1 \\ L & \text{if both players chose } (R, L) \text{ at } t - 1 \\ rand\{L, R\} & \text{Otherwise} \end{cases} \quad (3)$$

Reactive strategies also provide a way of maintaining coordination among a group of players. Grim trigger, for example, is a reactive strategy for the players in iterated prisoner's dilemma to maintain mutual cooperation. There exists a set of trigger strategies in a repeated game, by which the coordination among a group of players can be enforced. Once a group of players have coordinated their actions, they switch to the trigger strategy that one player will choose the minimax strategy if any other player in the group deviates from their coordination strategy.

## Type-*k* Equilibrium

One assumption in game theory is that the players believe that a deviation in their own strategy will not cause deviations by any other players. This is not a reasonable assumption in repeated games because of the existence of reactive strategies.

|  |  | Column Player | |
|---|---|---|---|
|  |  | $L$ | $R$ |
| Row Player | $L$ | 0, 0 | 1, 1 |
|  | $R$ | 1, 1 | 0, 0 |

**Fig 2. Payoff matrix of a coordination game.** Two players are indifferent between two pure-strategy NEs. The probability that any NE is achieved is 0.5 no matter what pure or mixed strategy they choose. Unless some reactive strategies are adopted, the probability of convergence to any NE is 0.5 in a repeated version of this game.

Coordination among a group of players can be achieved when they adopt specific reactive strategies, which may lead to equilibrium other than Nash equilibrium in repeated games.

**Definition 1**: In a repeated $n$-player game, a *type-k coordination* ($2 \leq k \leq n$) denotes that a group of $k$ players coordinate their actions by adopting some trigger strategies such that they will change their strategies simultaneously once any player in the group deviates from the assigned action.

The necessary condition of a type-$k$ coordination is that $k$ players can be better off by coordinating their actions. Let $v_i$ be the minimax payoff of player $i \in I$ and $s_i^*$ the minimax strategy. Let $K$ denote the group of $k$ players where $K \in I$. The necessary condition for a type-$k$ coordination is that there exists a strategy profile $\{\bar{s}_i\}$ ($i \in K$) such that

$$v_i < u_i(\bar{s}_1, \cdots, \bar{s}_k, \underbrace{s_{k+1}, \cdots s_n}_{n-k}) \tag{4}$$

hold for all $i \in K$ and whatever $\{s_j\}$ ($j \notin K$).

A type-$k$ coordination can be maintained if all players involved adopt a trigger strategy like this: keep playing the coordination strategy if all other players play their coordination strategies; otherwise, play the minimax strategy.

Given $\{\bar{s}_i\}$ ($i \in K$), the best responses of the players who do not belong to $K$ can be determined. Let $\{\bar{s}_i\}$ ($i \in I$) denote the strategy profile of all players. If $k$ players cannot further improve their payoffs by deviating from $\{\bar{s}_i\}$ simultaneously, the strategy profile $\{\bar{s}_i\}$ ($i \in I$) is a stable state (equilibrium) in the repeated game. This equilibrium is different from the concept of NE in that $k$ players coordinate their actions.

**Definition 2**: In an infinitely repeated $n$-player game, we call it a *type-k equilibrium* ($2 \leq k \leq n$) if a group of $k$ players coordinate their actions and they have no incentive to deviate from their strategies simultaneously.

A strategy profile $\{\bar{s}_i\}$ is a type-$k$ equilibrium if

$$u_i(s_1', \cdots, s_k', \underbrace{s_{k+1}, \cdots s_n}_{n-k}) \leq u_i(\bar{s}_1, \cdots, \bar{s}_k, \underbrace{s_{k+1}, \cdots s_n}_{n-k}) \tag{5}$$

are satisfied for any $\{s_i'\}$ ($i \in K$ and $s_i' \neq \bar{s}_i$) and $\{s_j\}$ ($j \notin K$).

We prove that any type-$k$ equilibrium is also a NE in the below proposition.

**Proposition 1**: In an infinitely repeated $n$-player game, any type-$k$ equilibrium ($2 \leq k \leq n$) is a NE.

Proof: Consider a strategy profile $\{\bar{s}_i\}$ that satisfies (5). Let's first consider the case of $k = n$. We have $v_i < u_i(\bar{s}_1, \ldots, \bar{s}_n)$ for all $i \in I$. According to the folk theorem, $\{\bar{s}_i\}$ is a NE.

In the case of $k < n$, if $v_i < u_i(\bar{s}_1, \ldots, \bar{s}_n)$ are satisfied for all $i \in I$, $\{\bar{s}_i\}$ is a NE according to the folk theorem. It is impossible that, for any player $i$, there is $v_i > u_i(\bar{s}_1, \ldots, \bar{s}_n)$ because player $i$ could deviate from $\bar{s}_i$ to the minimax strategy so that the payoff is guaranteed to be $v_i$. This conflicts with the fact that $\bar{s}_i$ is player $i's$ best response. We simply need to consider $v_i = u_i(\bar{s}_1, \ldots, \bar{s}_n)$ for some players $i \notin K$.

Let $M$ denote the group of players who receive their minimax payoffs. Any player $i \in M$ cannot improve his\her payoff by deviating from $\bar{s}_i$ since $\bar{s}_i$ is the best response to $\bar{s}_{-i}$.

Any player $i \in K$ cannot improve their payoff by deviating from $\bar{s}_i$. If player $i$ does deviate from $\bar{s}_i$ in order to gain a higher payoff in the current round, all other members of $K$ will play their minimax strategies in the future rounds. Player $i$ will have to play the minimax strategy and will receive $v_i$ in the future rounds. Knowing this, player $i$ has no incentive to deviate from $\bar{s}_i$.

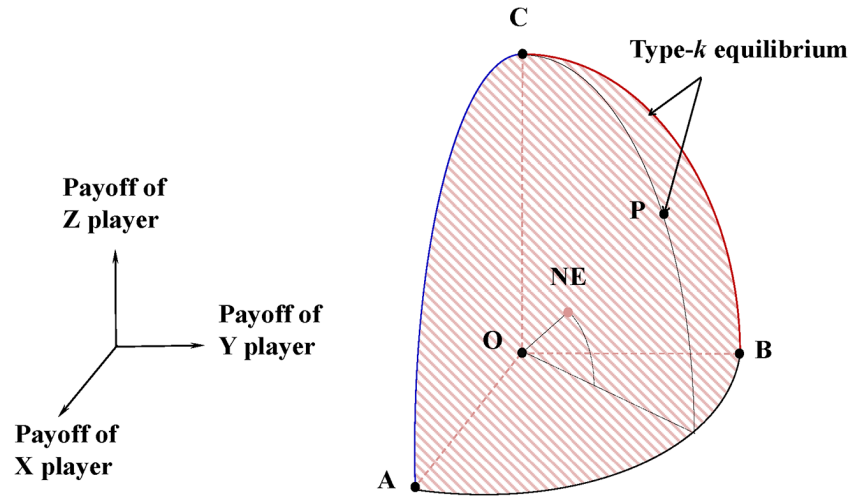

**Fig 3. The relationship between proposition 1 and the folk theorem in infinitely repeated three-player games.** O denotes the minimax payoff profile. Any feasible payoff profile within OABC (it includes the surface of ABC, but excludes the curves AB, BC, and CA.) is a NE according to the folk theorem. Propostion 1 proves that any feasible payoff profile on the ABC surface (including the curves AB, BC, and CA) can be NE if [5] holds. Any type-*k* equilibrium is a Parteto optimum.

Since any player has no incentive to deviate from $\{\bar{s}_i\}$, it is a NE.

Every type-*k* equilibrium is a NE and a NE is not necessarily a type-*k* equilibrium. Thus, the type-*k* equilibria are refinements of NE in repeated games.

The set of type-*k* equilibrium forms the Parteto frontier of all NEs in an infinitely repeated game. Any type-*k* equilibrium is a Pareto optimum for the group of *k* players. In three-player games, for example, the relationship between proposition 1 and the folk theorem can be illustrated by Fig 3.

For any group of players in an *n*-player game, there must be a strategy profile $\{\bar{s}_i\}$ such that these players cannot improve their payoffs by changing their strategies simultaneously. This strategy profile can be a type-*k* equilibrium if (4) and (5) are satisfied for some players. We prove the existence of type-*k* equilibrium in general repeated games in proposition 2.

**Proposition 2:** In an infinitely repeated *n*-player game where there exists two or more NE, there must be at least one type-*k* equilibrium.

Proof: When there exists two or more NE, there must be at least one strategy profile that is different from the minimax profile in a NE. Let $\{s_i\}$ denote such a strategy profile. We first prove that there must be $v_i < u_i(s_1, \ldots, s_n)$ for at least two players. Assume that there is $v_a < u_a(s_1, \ldots, s_n)$ for the player *a* and $v_i = u_i(s_1, \ldots, s_n)$ for any $i \neq a$. Since all players except *a* play their minimax strategies and they have no incentive to deviate unilaterally (because $\{s_i\}$ is a NE), $s_a$ is the minimax strategy for *a*. This conflicts with the premise that $\{s_i\}$ is different from the minimax profile. Thus, there must be $v_i < u_i(s_1, \ldots, s_n)$ for at least two players.

Suppose that there are $v_i < u_i(s_1, \ldots, s_n)$ for *k* players in the NE. If those *k* players cannot improve their payoffs by changing their strategies simultaneously, this NE is a type-*k* equilibrium. Otherwise, there must be a strategy profile $\{s'_i\}$ such that those *k* players cannot further improve their payoffs by changing their strategies simultaneously and $\{s'_i\}$ is a type-*k* equilibrium.

A NE is stable if a small change in the strategy of one player leads to a situation such that

a.  the player who did not change has no better strategy.

b.  the player who did change is now playing with a strictly worse strategy.

A type-*k* equilibrium is not stable if it is not a NE in the stage game because once a player within the coalition changes his/her strategy in a type-*k* equilibrium, all other *k*−1 players will be triggered to change their strategies. We do not concern the players excluded from the coalition because any change in their strategies has no influence on the coalition.

A type-*k* equilibrium is stable if it is also a NE in the stage game and the NE is stable. For example, the pure-strategy NEs in Fig 2 are type-2 equilibria and they are stable.

We have discussed the existence of type-*k* equilibrium in infinitely repeated games without discounting. In a repeated game with discounting, the discounted future payoffs must be greater than the excess current payoff due to deviating from the type-*k* equilibrium in order for each player in a type-*k* equilibrium to persist their strategies. Consider a constant discount factor $\delta \in (0,1)$ so that the summation of player *i*'s payoff in $T+1$ periods is $\sum_{t=0}^{T} \delta^t u_i^t$. Let $u_i'$ denote the maximum payoff of player *i* in the stage game given that *i* deviates from the type-*k* equilibrium while all players except *i* keep their strategies unchanged. For each player *i* within the coordination group, there should be

$$\lim_{T \to \infty} \sum_{t=1}^{T} \delta^t (u_i^t - v_i) > u_i' - u_i^0 \qquad (6)$$

This is the necessary condition for the existence of type-*k* equilibrium in infinitely repeated games with discounting.

## An Example

This example is to show the multiplicity of equilibria in repeated games. Consider a three-player game as shown in Fig 4. The option *R* is dominated by *L* for every player and the strategy profile (*L*, *L*, *L*) is the unique NE in the stage game.

There are numerous type-3 and type-2 equilibria in the infinitely repeated version of this game. A strategy profile is a type-3 equilibrium when X and Y choose (*R*, *R*) and Z chooses whatever mixed strategy in every round. There is a typical type-2 equilibrium when X and Y alternately choose (*L*, *R*) and (*R*, *L*) and Z chooses *L* (the point F in Fig 5).

If it is a repeated game with discounting, the necessary condition of the above type-2 equilibrium is that, for players X and Y, there are $\lim_{T \to \infty} \sum_{t=1}^{T} 3\delta^t > 1$, or $\delta > \frac{1}{4}$.

## Conclusions

Proposition 1 is a supplement to the folk theorem. The folk theorem proves the existence of the type-*n* equilibrium in repeated *n*-player games. Proposition 1 extends it to the general case of type-*k* equilibrium ($2 \le k \le n$).

If player **Z** chooses *L*:

|  |  | PLAYER **Y** | |
|---|---|---|---|
|  |  | *L* | *R* |
| PLAYER **X** | *L* | 2, 2, 2 | 9, 1, -1 |
|  | *R* | 1, 9, -1 | 3, 3, 4 |

If player **Z** chooses *R*:

|  |  | PLAYER **Y** | |
|---|---|---|---|
|  |  | *L* | *R* |
| PLAYER **X** | *L* | 2, 2, 1 | 7, 1, -2 |
|  | *R* | 1, 7, -2 | 4, 4, 3 |

**Fig 4. The payoff matrix of a three-player game is given, where three players are X, Y and Z and each has two options, *L* and *R*.**

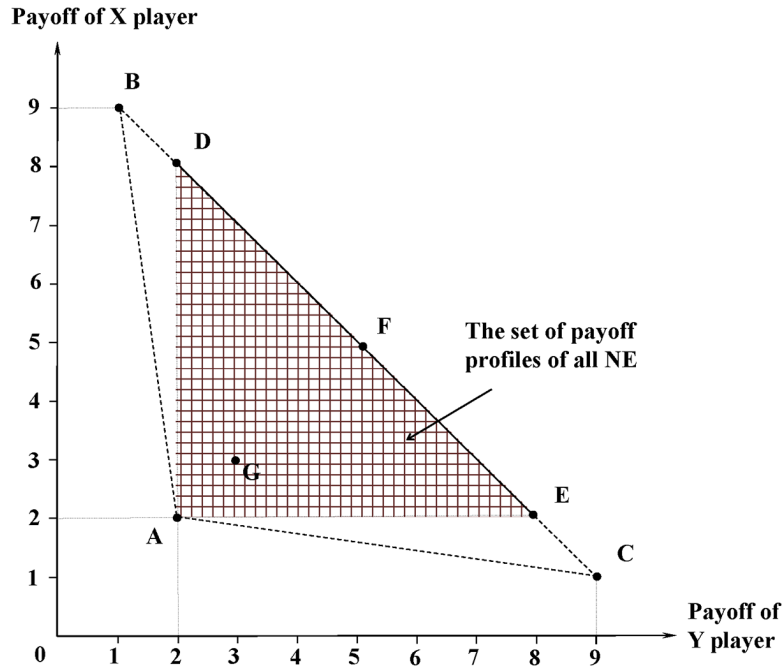

**Fig 5. The set of payoff profiles of all NE is a 3D polyhedron in the payoff space of X, Y, Z players and △ADE is its projection onto the X-Y plane.** △ABC is the projection of the set of feasible profiles. The point A represents the minimax profile. The point G represents a type-3 equilibrium. Any point on the segment DE represents a type-2 equilibrium.

Type-k equilibrium is a solution concept for repeated non-cooperative games. A type-*k* equilibrium is a Pareto optimum for the group of *k* players. In a type-*k* equilibrium not only does any individual player not have incentive to unilaterally change their strategies but also a group of *k* players has no incentive to deviate from it collectively, which means that the type-*k* equilibrium is stronger than NE in stability.

Type-*k* equilibrium is different from other refinements of NE, such as strong NE [20] and coalition proof NE [21, 22], in that a type-*k* equilibrium is not necessarily a NE in the stage game and it does not need communication or mediation among players. Type-*k* equilibrium is different from the concepts of coalition [23,24] or core [25] of cooperative games in that the players in a type-*k* equilibrium are payoff-maximized and they make their choices independently. A type-*k* equilibrium does not exist in one-shot game because coordination among the players can be formed only if all players choose to adopt specific reactive strategies. Reactive strategies are strategies for repeated games and they are neither pure strategies nor mixed strategies in the stage game.

The emergence of cooperation in evolutionary dynamics has attracted a great deal of research [26–36]. The type-*k* equilibrium suggests that cooperation in evolution starts from a group of players rather than an individual player. A group of players would coordinate their actions by adopting specific reactive strategies if they could be better off by doing so. This collective behaviour is much more effective and robust than any individual behaviour in building and maintaining cooperation in evolution.

Backward induction has led to some controversy despite of its wide use in finitely repeated games [37, 38]. It is obvious that the choices of players cannot be backward inducted when some of them adopt specific reactive strategies. In finitely repeated games, the end game effect cannot prevent the players from adopting specific reactive strategies. When both past actions and the end game effect have an influence on the strategies of players, transition from one NE to another is possible.

The existence of type-*k* equilibrium explains, to some extent, why the biodiversity in evolutionary games is much more complex than classical game theory has predicted. The type-*k* equilibrium belongs to the set of NE that has been neglected in non-cooperative game theory. This set of NE possibly contains more complicated equilibrium than the type-*k* equilibrium, for example the equilibrium that has two or more coalitions in it.

## Author Contributions

Conceived and designed the experiments: JL. Performed the experiments: JL. Analyzed the data: JL. Wrote the paper: JL GK.

## References

1. Maynard Smith J. (1982) Evolution and the Theory of Games. Cambridge University Press. ISBN 0-521-28884-3.

2. Cressman R. (2003) Evolutionary Dynamics and Extensive Form Games (Vol. 5). The MIT Press.

3. Cressman R. and Tao Y. (2014) The replicator equation and other game dynamics, Proceedings of the National Academy of Sciences 111: 10810–10817.

4. Friedman J. (1971) A non-cooperative equilibrium for supergames. Review of Economic Studies 38(1): 1–12.

5. Rubinstein A. (1979) Equilibrium in Supergames with the Overtaking Criterion, Journal of Economic Theory, 21: 1–9.

6. Fudenberg D. and Maskin E. (1986) The folk theorem in repeated games with discounting or with incomplete information. Econometrica: Journal of the Econometric Society: 533–554.

7. Ely J. and Välimäki J. (2002) A robust folk theorem for the prisoner's dilemma. Journal of Economic Theory 102(1): 84–105.

8. Dutta P. (1995) A folk theorem for stochastic games. Journal of Economic Theory, 66(1): 1–32.

9. Fudenberg D. and Levine D. (1991) An approximate folk theorem with imperfect private information. Journal of Economic Theory, 54(1): 26–47.

10. Wen Q. (2002) A folk theorem for repeated sequential games. The Review of Economic Studies, 69(2): 493–512.

11. Abreu D., Dutta P. K., & Smith L. (1994) The folk theorem for repeated games: a NEU condition. Econometrica: Journal of the Econometric Society, 939–948.

12. Baye M. R. and Morgan J. (1999) A folk theorem for one-shot Bertrand games.Economics Letters, 65(1), 59–65.

13. Hörner J. and Olszewski W. (2006) The Folk Theorem for Games with Private Almost-Perfect Monitoring. Econometrica, 74(6), 1499–1544.

14. Matsushima H. (1991) On the theory of repeated games with private information: Part I: anti-folk theorem without communication. Economics Letters, 35(3), 253–256.

15. Obara I. (2009) Folk theorem with communication. Journal of Economic Theory, 144(1), 120–134.

16. Li J., Hingston P. and Kendall G. (2011) Engineering design of strategies for winning iterated prisoner's dilemma competitions. IEEE Transactions on Computational Intelligence & AI in Games, 3(4): 348–360.

17. Press W. and Freeman D. (2012) Iterated Prisoner's Dilemma contains strategies that dominate any evolutionary opponent. Proceedings of the National Academy of Sciences 109(26): 10409–10413.

18. Hilbe C., Nowak M., and Sigmund K. (2013) The evolution of extortion in iterated Prisoner's Dilemma games. Proceedings of the National Academy of Sciences 110(17): 6913–6918.

19. Stewart A. and Plotkin J. (2012) Extortion and cooperation in the Prisoner's Dilemma. Proceedings of the National Academy of Sciences 109(26): 10134–10135.

20. Aumann R. (1959) Acceptable points in general cooperative n-person games. in Contributions to the Theory of Games IV, Princeton University Press, Princeton, N.J.

21. Moreno D. and Wooders J. (1996) Coalition-Proof Equilibrium, Games and Economic Beahavior, 17: 82–112.

22. Mertens J. (2003) Ordinality in Non Cooperative Games, International Journal of Game Theory, 32: 387–430.

23. Olson M. (2009) The logic of collective action. (Vol. 124) Harvard University Press.

24. Shenoy P. (1979). On coalition formation: a game-theoretical approach. International Journal of Game Theory, 8(3), 133–164.

25. Aumann, R. and Shapley, L. (1970). *Values of Non-atomic Games, IV: The Value and the Core* (No. RM-6260). RAND CORP SANTA MONICA CALIF.

26. Ferrierre R. and Michod R. (1995) Invading wave of cooperation in a spatial iterated prisoner's dilemma, Proceedings of the Royal Society B, 259(1354): 77–83.

27. Nowak M., Sasaki A., Taylor C. and Fudenberg D. (2004) Emergence of cooperation and evolutionary stability in finite populations. Nature, 428: 646–650.

28. Krams I., Kokko H., Vrublevska J., Abolins-Abols M., Krama T. and Rantala M. (2013) The excuse principle can maintain cooperation through forgivable defection in the prisoner's dilemma game, Proceedings of the Royal Society B, 280(1766): 20131475.

29. Li J. and Kendall G. (2013) Evolutionary stability of discriminating behaviors with the presence of kin cheaters, IEEE Transactions on Cybernetics, 43(6): 2044–2053.

30. Hauert C. and Schuster H. (1997) Effect of increasing the number of players and memory size in the iterated prisoner's dilemma: a numerical approach, Proceedings of the Royal Society B, 264: 513–519.

31. Killingback T, Doebeli M, Knowlton N (1999) Variable investment, the continuous prisoner's dilemma, and the origin of cooperation. Proceedings of the Royal Society of London Series B: Biological Sciences, 266: 1723–1728.

32. Perc M., & Wang Z. (2010) Heterogeneous aspirations promote cooperation in the prisoner's dilemma game. PLoS One, 5(12): e15117.

33. Meloni S., Buscarino A., Fortuna L., Frasca M., Gómez-Gardeñes J., Latora V., et al. (2009) Effects of mobility in a population of prisoner's dilemma players. Physical Review E, 79(6): 067101.

34. Poncela J., Gómez-Gardeñes J., Floría L. M., Sánchez A., and Moreno Y. (2008) Complex cooperative networks from evolutionary preferential attachment. PLoS one, 3(6): e2449.

35. Santos F. C., Santos M. D., and Pacheco J. M. (2008) Social diversity promotes the emergence of cooperation in public goods games. Nature, 454(7201): 213–216.

36. Braun D. A., Ortega P. A., and Wolpert D. M. (2009) Nash equilibria in multi-agent motor interactions. PLoS computational biology, 5(8): e1000468.

37. Li J., Kendall G. and Vasilakos A. (2013) Backward induction and repeated games with strategy constraints: an inspiration from the surprise exam paradox, IEEE Transactions on Computational Intelligence and AI in Games, 5(3): 242–250.

38. McCabe K. and Smith V. (2000) A comparison of naïve and sophisticated subject behaviour with game theoretic predictions, Proceedings of the National Academy of Sciences, 97(7): 3777–3781.